\documentclass{article}
\usepackage{amsfonts}
\usepackage{amsmath}

\input{tcilatex}
\begin{document}

\title{\textbf{Loop Quantum Gravity}\\
\textbf{in the Momentum Representation}}
\author{W. F. Chagas-Filho \\
Physics Department, Federal University of Sergipe, Brazil}
\maketitle

\begin{abstract}
We present a generalization of the first-order formalism used to describe
the dynamics of a classical system. The generalization is then applied to
the first-order action that describes General Relativity. As a result we
obtain equations that can be interpreted as describing quantum gravity in
the momentum representation.
\end{abstract}

\section{Introduction}

As is well-known, Quantum Mechanics can be formulated in the configuration
(or position) representation or in the momentum representation. This
situation emerges from the possible representations of the fundamental
commutators of the quantum theory. To illustrate this, consider the simple
example of the quantization of a one-dimensional system with a configuration
variable $q$ and a momentum variable $p$. The corresponding quantum
operators $\hat{q}$ and $\hat{p}$ must provide a representation of the
fundamental commutator 
\begin{equation}
\lbrack \hat{q},\hat{p}]=[\hat{q}\hat{p}-\hat{p}\hat{q}]=i\hbar  \tag{1}
\end{equation}%
The usual way to represent the commutator (1) is to choose 
\begin{equation}
\hat{q}=q\text{ \ \ \ \ \ \ \ \ \ \ \ \ }\hat{p}=-i\hbar \frac{d}{dq} 
\tag{2}
\end{equation}%
In this case the wave function will be a function of $q$, that is $\psi
=\psi (q)$, and we will be in the configuration representation.

Another possibility of representing the commutator (1) is to choose 
\begin{equation}
\hat{q}=i\hbar \frac{d}{dp}\text{ \ \ \ \ \ \ \ \ \ \ \ \ }\hat{p}=p  \tag{3}
\end{equation}%
In this case the wave function will be a function of $p$, that is $\psi
=\psi (p)$, and we will be in the momentum representation. From a naive
perspective, the operators (3) can be obtained from the operators (2) simply
by substituting the letter $q$ by $p$ and the letter $p$ by $-q$ in
equations (2). However, in a deeper level these two possibilities are
related to the quantum mechanical wave-particle duality. The configuration
representation is related to the particle aspect. Because of the De Broglie%
\'{}%
s relation $\lambda =h/p$, the momentum representation is related \ to the
wave aspect. The quantum wave-particle duality has a trace in classical
mechanics in the form of a Hamiltonian duality. This duality interchanges
position and momentum and leaves invariant the definition of the Poisson
bracket. In this paper we will use this classical Hamiltonian duality to
construct a formulation of quantum gravity in the momentum representation.

At present time, a quantum theory for the gravitational interaction, based
on the canonical quantization of General Relativity (GR) is under
development. It is called Loop Quantum Gravity. This theory has produced
interesting results, such as the quantization of area and volume in terms of
the Planck length $L_{P}=\sqrt{\frac{\hbar G}{c^{3}}}=1,62\times 10^{-35}m$.
But with no present available way to test the theory against experimental
results, the validity of LQG remains an open question [1,2,3].

In this paper we present a calculation that points in the direction of the
validity of LQG. Here we present the basic equations of a momentum space
formulation of quantum gravity, taking as the starting point the first-order
action for GR.

This paper is organized as follows. In section two we derive the two simple
classical equations that allow transitions to quantum gravity in the
configuration and in the momentum representations. In section three we
review the basic equations of quantum gravity in the configuration
representation (Loop Quantum Gravity). In section four we present the basic
equations of quantum gravity in the momentum representation. Brief
concluding remarks appear in section five.

\section{The first-order formalism and the transition to quantum mechanics}

The first-order formalism is in the interface between Lagrangian mechanics
and Hamiltonian mechanics. According to Dirac [6], a Hamiltonian formalism
is a first approximation to a corresponding quantum theory. Since quantum
mechanics can be formulated in the configuration or in the momentum
representations, we need two first-order formalisms, one for each
representation of quantum mechanics.

\subsection{The first-order formalism for the configuration space
formulation of quantum mechanics}

A first-order formalism which can be considered as the classical limit of a
configuration space formulation of quantum mechanics is the usual
first-order formalism. It is based on the action functional

\begin{equation}
S=\int_{t_{1}}^{t_{2}}dt[p\dot{q}-H(q,p)]  \tag{4}
\end{equation}%
where $H(q,p)$ is the Hamilton%
\'{}%
s function. Varying action (4) we find 
\begin{equation}
\delta S=\int_{t_{1}}^{t_{2}}dt[-\frac{\partial H}{\partial q}\delta
q+p\delta \dot{q}+(\dot{q}-\frac{\partial H}{\partial p})\delta p]  \tag{5}
\end{equation}%
Integrating by parts the second term we have%
\begin{equation*}
\int_{t_{1}}^{t_{2}}dtp\delta \dot{q}=p\delta q\mid
_{t_{1}}^{t_{2}}-\int_{t_{1}}^{t_{2}}dt\dot{p}\delta q
\end{equation*}%
Inserting this result into the variation (5) we are left with 
\begin{equation*}
\delta S=\int_{t_{1}}^{t_{2}}dt[-(\dot{p}+\frac{\partial H}{\partial q}%
)\delta q+(\dot{q}-\frac{\partial H}{\partial p})\delta p]+p\delta q\mid
_{t_{1}}^{t_{2}}
\end{equation*}%
The above integral vanishes if Hamilton%
\'{}%
s equations%
\begin{equation}
\dot{q}=\frac{\partial H}{\partial p}\text{ \ \ \ \ \ \ \ \ \ \ \ \ \ }\dot{p%
}=-\frac{\partial H}{\partial q}  \tag{6}
\end{equation}%
are satisfied. In this case the variation (5) reduces to the surface term 
\begin{equation*}
\delta S=p\delta q\mid _{t_{1}}^{t_{2}}
\end{equation*}%
Now we require that $\delta q(t_{1})=0$ and leave $\delta q$ arbitrary at $%
t=t_{2}$. We therefore see that, as a function of the final point of the
trajectory, action (4) satisfies 
\begin{equation}
p=\frac{\delta S}{\delta q}  \tag{7}
\end{equation}%
As we shall see below, equation (7) plays a central role in the transition
to quantum gravity in configuration space.

\subsection{The first-order formalism for the momentum space formulation of
quantum mechanics}

We now introduce a first-order formalism which can be considered as the
classical limit for a momentum space formulation of quantum mechanics. This
formalism can be constructed using the Hamiltonian duality transformation 
\begin{equation}
q\rightarrow p\text{ \ \ \ \ \ \ \ \ \ \ \ \ \ \ }p\rightarrow -q  \tag{8}
\end{equation}%
which leaves invariant the structure of the Hamilton%
\'{}%
s equations (6) and the definition of the Poisson bracket 
\begin{equation*}
\{A,B\}=\frac{\partial A}{\partial q}\frac{\partial B}{\partial p}-\frac{%
\partial A}{\partial p}\frac{\partial B}{\partial q}
\end{equation*}%
which defines the algebraic structure in the phase space $(q,p)$.

Applying the duality transformation (8) to action (4) we obtain the new
action {\Large \ }%
\begin{equation}
S=\int_{t_{1}}^{t_{2}}dt[-q\dot{p}-\tilde{H}(q,p)]  \tag{9}
\end{equation}

Varying action (9) we have {\Large \ }%
\begin{equation}
\delta S=\int_{t_{1}}^{t_{2}}dt[-(\dot{p}+\frac{\partial \tilde{H}}{\partial
q})\delta q-\frac{\partial \tilde{H}}{\partial p}\delta p-q\delta \dot{p}] 
\tag{10}
\end{equation}%
Integrating by parts the last term gives 
\begin{equation*}
-\int_{t_{1}}^{t_{2}}dtq\delta \dot{p}=-q\delta p\mid
_{t_{1}}^{t_{2}}+\int_{t_{1}}^{t_{2}}dt\dot{q}\delta p
\end{equation*}%
Inserting this result into the variation (10) we find that 
\begin{equation}
\delta S=\int_{t_{1}}^{t_{2}}dt[-(\dot{p}+\frac{\partial \tilde{H}}{\partial
q})\delta q+(\dot{q}-\frac{\partial \tilde{H}}{\partial p})\delta p]-q\delta
p\mid _{t_{1}}^{t_{2}}  \tag{11}
\end{equation}%
Now, if Hamilton%
\'{}%
s equations 
\begin{equation*}
\dot{q}=\frac{\partial \tilde{H}}{\partial p}\text{ \ \ \ \ \ \ \ \ }\dot{p}%
=-\frac{\partial \tilde{H}}{\partial q}
\end{equation*}%
are valid, the variation (11) reduces to the surface term 
\begin{equation*}
\delta S=-q\delta p\mid _{t_{1}}^{t_{2}}
\end{equation*}%
We now impose that $\delta p(t_{1})=0$ and leave $\delta p$ arbitrary at $%
t=t_{2}$. We then see that, as a function of the end point, action (9)
satisfies 
\begin{equation}
-q=\frac{\delta S}{\delta p}  \tag{12}
\end{equation}%
Equation (12) is the central equation in this paper. As we will see below,
it allows the the transition to quantum gravity in momentum space.

\subsection{The transition to quantum mechanics}

It is important to stress that the first-order formalism of section 2.2 was
introduced to be used as the classical limit of a momentum space formulation
of quantum mechanics. Since classical mechanics can not be formulated in
momentum space because the quantum wave-particle duality practically
disappears at the classical level, the classical Hamilton equations for the
dynamic variables $q$ and $p$ derived from the Hamiltonian $\tilde{H}(q,p)$
will in general appear to be inconsistent. However, the quantum Schr\"{o}%
dinger equation obtained from the quantum operator corresponding to $\tilde{H%
}(q,p)$ will be consistent.

The simplest example of the above situation is a free non-relativistic
particle, described by the Hamiltonian%
\begin{equation*}
H=\frac{p^{2}}{2m}
\end{equation*}%
The Hamilton equations for this system are%
\begin{equation*}
\dot{q}=\frac{\partial H}{\partial p}=\frac{p}{m}\text{ \ \ \ \ \ \ \ \ \ \
\ \ \ }\dot{p}=-\frac{\partial H}{dq}=0
\end{equation*}%
Quantization of this system using the operators (2) leads to the Schr\"{o}%
dinger equation%
\begin{equation*}
-\text{%
h{\hskip-.2em}\llap{\protect\rule[1.1ex]{.325em}{.1ex}}{\hskip.2em}%
}^{2}\frac{\partial ^{2}\psi (q,t)}{\partial q^{2}}=i\text{%
h{\hskip-.2em}\llap{\protect\rule[1.1ex]{.325em}{.1ex}}{\hskip.2em}%
}\frac{\partial \psi (q,t)}{\partial t}
\end{equation*}%
Now, from the duality transformation (8), we obtain the Hamiltonian%
\begin{equation*}
\tilde{H}=\frac{q^{2}}{2m}
\end{equation*}%
The Hamilton equations now are%
\begin{equation*}
\dot{q}=\frac{\partial \tilde{H}}{\partial p}=0\text{ \ \ \ \ \ \ \ \ \ \ \
\ \ }\dot{p}=-\frac{\partial \tilde{H}}{\partial q}=-\frac{q}{m}
\end{equation*}%
which look inconsistent because the particle does not move while its
momentum varies. However, quantization of this system using the operators
(3) leads to the Schr\"{o}dinger equation 
\begin{equation*}
-\text{%
h{\hskip-.2em}\llap{\protect\rule[1.1ex]{.325em}{.1ex}}{\hskip.2em}%
}^{2}\frac{\partial ^{2}\psi (p,t)}{\partial p^{2}}=i\text{%
h{\hskip-.2em}\llap{\protect\rule[1.1ex]{.325em}{.1ex}}{\hskip.2em}%
}\frac{\partial \psi (p,t)}{\partial t}
\end{equation*}%
which is perfectly consistent.

\section{Loop Quantum Gravity}

In this section we review the basic equations that define quantum gravity in
the configuration representation (LQG).

In 1986 Ashtekar [4,5] introduced a new set of variables to describe General
Relativity. In this new set of variables GR can be described by the
first-order action ( for details on this construction see ref. [1], chapter
four ) 
\begin{equation}
S=\frac{1}{8\pi iG}\int d^{4}x(E_{i}^{a}\dot{A}_{a}^{i}-\lambda
^{i}D_{a}E_{i}^{a}-\lambda ^{a}F_{ab}^{i}E_{i}^{b}-\lambda
F_{ab}^{ij}E_{i}^{a}E_{j}^{b})  \tag{13}
\end{equation}%
where{\Large \ }%
\begin{equation*}
D_{a}V^{i}=\partial _{a}V^{i}+\epsilon _{jk}^{i}A_{a}^{j}V^{k}
\end{equation*}%
is the covariant derivative, 
\begin{equation*}
F_{ab}^{i}=\partial _{a}A_{b}^{i}-\partial _{b}A_{a}^{i}+\epsilon
_{jk}^{i}A_{a}^{j}A_{b}^{k}
\end{equation*}%
is the curvature and $F_{ab}^{ij}=\epsilon _{k}^{ij}F_{ab}^{k}$. The
variables $\lambda ^{i}$, $\lambda ^{a}$ and $\lambda $ are Lagrange
multipliers without dynamics.

Indices $i,j,...=1,2,3$ are internal $SU(2)$ indices and $a,b=1,2,3$ are
space indices. Comparing action (13) with action (4) we see that

a) the configuration variable is $A_{a}^{i}(\vec{x})$

b) the canonical momentum is{\Large \ }$E_{i}^{a}(\vec{x})$

c) the total [6] Hamiltonian density is given by $H_{T}=\lambda
^{i}D_{a}E_{i}^{a}+\lambda ^{a}F_{ab}^{i}E_{i}^{b}+\lambda
F_{ab}^{ij}E_{i}^{a}E_{j}^{b}${\Large \ }

\bigskip

Varying action (13) in relation to the variables $\lambda ^{i},$ $\lambda
^{a}$ and{\Large \ }$\lambda $ we obtain the first-class [6] constraints 
\begin{equation}
D_{a}E_{i}^{a}=0  \tag{14a}
\end{equation}%
\begin{equation}
F_{ab}^{i}E_{i}^{b}=0  \tag{14b}
\end{equation}%
\begin{equation}
F_{ab}^{ij}E_{i}^{a}E_{j}^{b}=0  \tag{14c}
\end{equation}%
Equation (14a) is the requirement of invariance of the theory under internal 
$SU(2)$ transformations. Equation (14b) is the requirement of invariance of
the theory under space diffeomorphisms. Equation (14c) is the canonical
Hamiltonian. Equations (14) are equivalent to the Einstein equations [1]

Now, using the analog of equation (7), that is {\Large \ }%
\begin{equation*}
E_{i}^{a}=\frac{\delta S}{\delta A_{a}^{i}}
\end{equation*}%
we make a transition to the Hamilton-Jacobi formalism as an intermediate
step to the quantum theory. Equations (14) then become [1] 
\begin{equation}
D_{a}\frac{\delta S}{\delta A_{a}^{i}}=0  \tag{15a}
\end{equation}%
\begin{equation}
F_{ab}^{i}\frac{\delta S}{\delta A_{b}^{i}}=0  \tag{15b}
\end{equation}%
\begin{equation}
F_{ab}^{ij}\frac{\delta S}{\delta A_{a}^{i}}\frac{\delta S}{\delta A_{b}^{j}}%
=0  \tag{15c}
\end{equation}%
The transition to the quantum theory in the configuration space is then
obtained by substituting the classical action {\Large \ }$S$ by the wave
functional $\Psi (A)$ in equations (15). The final result is{\Large \ }%
\begin{equation}
D_{a}\frac{\delta }{\delta A_{a}^{i}}\Psi (A)=0  \tag{16a}
\end{equation}%
\begin{equation}
F_{ab}^{i}\frac{\delta }{\delta A_{b}^{i}}\Psi (A)=0  \tag{16b}
\end{equation}%
\begin{equation}
F_{ab}^{ij}\frac{\delta }{\delta A_{a}^{i}}\frac{\delta }{\delta A_{b}^{j}}%
\Psi (A)=0  \tag{16c}
\end{equation}%
Equations (16) are the quantum gravity equations [1] in the configuration
representation.

\section{Quantum gravity in the momentum representation}

In this section we derive the basic equations that we interpret to define
quantum gravity in the momentum representation. First we apply the duality
transformation {\Large \ }%
\begin{equation}
A_{a}^{i}\rightarrow E_{i}^{a}\text{ \ \ \ \ \ \ \ \ \ \ \ \ \ \ }%
E_{i}^{a}\rightarrow -A_{a}^{i}  \tag{17}
\end{equation}%
to the first-order action (13) for GR. We obtain the action{\Large \ }%
\begin{equation}
S=\frac{1}{8\pi iG}\int d^{4}x(-A_{a}^{i}\dot{E}_{i}^{a}-\lambda _{i}\nabla
^{a}A_{a}^{i}-\lambda _{b}R_{i}^{ab}A_{a}^{i}-R_{ij}^{ab}A_{a}^{i}A_{b}^{j})
\tag{18}
\end{equation}%
where the covariant derivative $\nabla ^{a}$ is defined by {\Large \ }%
\begin{equation*}
\nabla ^{a}V^{i}=\partial ^{a}V^{i}+\epsilon ^{ijk}E_{j}^{a}V^{k}
\end{equation*}%
and the curvature $R_{i}^{ab}$ is given by%
\begin{equation*}
R_{i}^{ab}=\partial ^{a}E_{i}^{b}-\partial ^{b}E_{i}^{a}+\epsilon
_{i}^{jk}E_{j}^{a}E_{k}^{b}
\end{equation*}%
with $R_{ij}^{ab}=\epsilon _{ijk}R_{k}^{ab}$. It is important to mention
here that we should not interpret action (18) as describing a classical
physical system. Rather, it should be interpreted as a formal equation
describing the classical limit of a \textsl{quantum }physical system.

The equations of motion for the variables $\lambda _{i}$, $\lambda _{a}$ and%
{\Large \ }$\lambda $ give the first-class [6] constraints{\Large \ }%
\begin{equation}
\nabla ^{a}A_{a}^{i}=0  \tag{19a}
\end{equation}%
\begin{equation}
R_{i}^{ab}A_{b}^{i}=0  \tag{19b}
\end{equation}%
\begin{equation}
R_{ij}^{ab}A_{a}^{i}A_{b}^{j}=0  \tag{19c}
\end{equation}%
The next step towards the quantum theory is to use the general equation (9)
we derived in section two. In the present case equation (12) becomes{\Large %
\ }%
\begin{equation}
A_{a}^{i}=-\frac{\delta S}{\delta E_{i}^{a}}  \tag{20}
\end{equation}%
Substituting equation (20) into equations (19) we have 
\begin{equation}
\nabla ^{a}\frac{\delta S}{\delta E_{i}^{a}}=0  \tag{21a}
\end{equation}%
\begin{equation}
R_{i}^{ab}\frac{\delta S}{\delta E_{i}^{b}}=0  \tag{21b}
\end{equation}%
\begin{equation}
R_{ij}^{ab}\frac{\delta S}{\delta E_{i}^{a}}\frac{\delta S}{\delta E_{j}^{b}}%
=0  \tag{21c}
\end{equation}%
Finally, the transition to the quantum theory is completed by substituting
the classical action{\Large \ }$S$ by the wave functional $\Psi (E)$ in
momentum space. This gives {\Large \ }%
\begin{equation}
\nabla ^{a}\frac{\delta }{\delta E_{i}^{a}}\Psi (E)=0  \tag{22a}
\end{equation}%
\begin{equation}
R_{i}^{ab}\frac{\delta }{\delta E_{i}^{b}}\Psi (E)=0  \tag{22b}
\end{equation}%
\begin{equation}
R_{ij}^{ab}\frac{\delta }{\delta E_{i}^{a}}\frac{\delta }{\delta E_{j}^{b}}%
\Psi (E)=0  \tag{22c}
\end{equation}%
We interpret equations (22) as the quantum gravity equations in the momentum
representation.

\section{Conclusions}

In this paper we presented a generalization of the first-order formalism
used to describe the dynamics of a classical system. This generalization is
based on the Hamiltonian duality that interchanges the configuration and the
momentum variables. The generalization is then applied to the first-order
action that describes General Relativity. As a result of this, we obtain
equations that can be interpreted as describing quantum gravity in the
momentum representation. The conclusion of this paper is that, as quantum
mechanics, quantum gravity can be formulated in the configuration or in the
momentum representation.

\bigskip

\bigskip

{\Large \ \ }


\begin{thebibliography}{9}
\bibitem{1} C. Rovelli, \textsl{Quantum Gravity}, Cambridge Monographs on
Mathematical Physics, 2004

\bibitem{2} T. Thiemann, \textsl{Modern Canonical Quantum General Relativity,%
} Cambridge Monographs on Mathematical Physics, 2008

\bibitem{3} C. Rovelli and F. Vidotto, \textsl{Covariant Loop Quantum Gravity%
}, Cambridge University Press, 2015

\bibitem{4} A. Ashtekar, \textsl{New variables for classical and quantum
gravity}, Phys. Rev. Lett. 57 (1986) 2244

\bibitem{5} A. Ashtekar, \textsl{New Hamiltonian formulation of general
relativity}, Phys. Rev. D36 (1987) 1587

\bibitem{6} P. A. M. Dirac, \textsl{Lectures on Quantum Mechanics}, Yeshiva
University, 1964
\end{thebibliography}
\end{document}